# Dispersion interaction between two bodies of arbitrary shape

M.V. Davidovich

Saratov National Research State University

410012 Saratov, Russia,

davidovichmv@yandex,ru

*Abstract* – Based on the vacuum tensor Green's functions of electrodynamics, the Lippmann-Schwinger-type equations were derived for the correlations of electromagnetic fields. These correlations were applied within the Maxwell stress tensor and the Lorentz force formula to determine Casimir dispersion forces for two arbitrary bodies. The analysis is performed by the excitation of bodies by electric and magnetic dipoles.

## 1. Introduction

Although the Casimir forces or dispersion forces for macro objects were introduced about eighty years ago [1], the main results were obtained mainly for plane–layered structures [2-44], the Lifshitz formula for half-spaces [2-6] and the van Kampen-Schram form formulas [9,12,13] for plane-layered structures are most often used for this. Van Kampen-Schram form is the generalization of formula [2]. For bodies with a curved surface, approximations based on the formula [2] (the so-called proximity force applications [45]) are used. The difficulties of the classical description are related to the need to construct a tensor Green's functions (GFs) of structure satisfying the equation with discontinuous coefficients and boundary conditions [42]. The GF of a plane-layered structure is relatively simple to construct [8,19]. Description of dispersion interactions is possible within the framework of macroscopic quantum electrodynamics [3-6,17,18,38,41]. However, it is quite difficult to implement for bodies of arbitrary shape, and so far no results have been obtained for them. There are several papers describing the dispersion interactions of bodies of complex and even arbitrary shapes, for example, [39, 46-50], but the results are given for cylinders and balls. They are based on a number of approaches: perturbation theory, the theory of matrix representations of fields using scattering matrices, etc., which often also contain approximations. Methods of multiple scattering formalism are being developed for bodies of arbitrary shape [49,50]. Methods of nonequilibrium electromagnetic fluctuations and dispersion interactions are also being developed [37,39,43,51,52,53].

In this paper, we consider a new approach based on the representation of fields in terms of Green's vacuum functions (GFs) and the Maxwell stress tensor (MST). An approach based on the Lorentz force is also considered. In addition, the correlation relations (CR) obtained by exciting two bodies to dipoles are considered. The use of the Lorentz force seems to be preferable, since, unlike MST, it is determined in any medium where the magnetic induction vector **B** can be introduced. The equivalence of the approach based on the Lorentz formula and MST follows from the theorem (law) on the conservation of momentum of the electromagnetic field and matter [54,55]. The sources that create the field transmit an impulse to it, receiving a return. The field pulse acts on bodies, i.e. it is transmitted to them in the form of reflection of electromagnetic waves and in the form of their absorption in the volume of bodies. Considering the fluctuation sources (densities of polarization currents) of the second body as sources of its field, one can find its interaction with the first body. To do this, it is sufficient to consider the polarization currents of the first body and the volume and surface charge densities induced in it. For simplicity, we consider bodies to be homogeneous and do not consider bulk charge densities. We also believe that the body's own fields do not act on the polarization currents and charges that arise in it (there is no self-action). Then the force will depend on the distance between the bodies and their shape Fig. 1. As a result, fields can be expressed in terms of fluctuation current densities, and field correlations can be expressed in terms of correlations of these current densities. To determine correlations of fluctuations of currents or fields, the Lippmann-Schwinger integral equation (LSIE) should be solved. By formulating a problem based on tensor GF pf structures, it is possible to formulate a Dyson-type equation for such a GF.

## 2. Problem statement and integral equations

Let us consider two macroscopic bodies described by macroscopic permittivity (DP) $\hat{\varepsilon}(\omega,\mathbf{r})$ and magnetic permeability (MP) $\hat{\mu}(\omega,\mathbf{r})$. In general, they are tensor, and the bodies are anisotropic. At least two bodies $V=V_1+V_2$ must be considered for the introduction of force. In reality, bodies at high frequencies do not possess MP, but are described by a tensor DP with spatial dispersion (SD). In the case of a single body, the interaction of its parts can be considered. For simplicity and convenience, we will first limit ourselves to constant values of permeabilities in bodies, i.e. we will use permeabilities inside bodies $\hat{\varepsilon}_1(\omega)$, $\hat{\varepsilon}_2(\omega)$, $\hat{\mu}_1(\omega)$, $\hat{\mu}_2(\omega)$.. DP is described by one symbol $\hat{\varepsilon}(\omega)$, considering it to be different and constant in bodies. As you will see, this does not limit the generality in any way. The heterogeneity of bodies only leads to additional volume charges. Also, for simplicity, we will consider bodies to be isotropic. Anisotropic properties are generalized naturally. Next, we use the inverse Fourier transform to express the components of the fields in terms of their spectral functions, for example,

$$E_\alpha(t,\mathbf{r}) = \frac{1}{2\pi}\int_{-\infty}^{\infty} E_\alpha(\omega,\mathbf{r})\exp(i\omega t)d\omega = \\ = \frac{1}{\pi}\int_{0}^{\infty} E_\alpha(\omega,\mathbf{r})\exp(i\omega t)d\omega \quad . \tag{1}$$

Here $E_\alpha(-\omega,\mathbf{r}) = E_\alpha^*(-\omega,\mathbf{r})$, $\alpha = x, y, z$. A macroscopic body in the **E**, **H** field is equivalent to the action of the densities of the electric current of polarization $\mathbf{J} = i\omega\varepsilon_0\left(\hat{\varepsilon} - \hat{I}\right)\mathbf{E}$ and the magnetic current of polarization $\mathbf{J}^\mu = i\omega\mu_0\left(\hat{\mu} - \hat{I}\right)\mathbf{H}$.

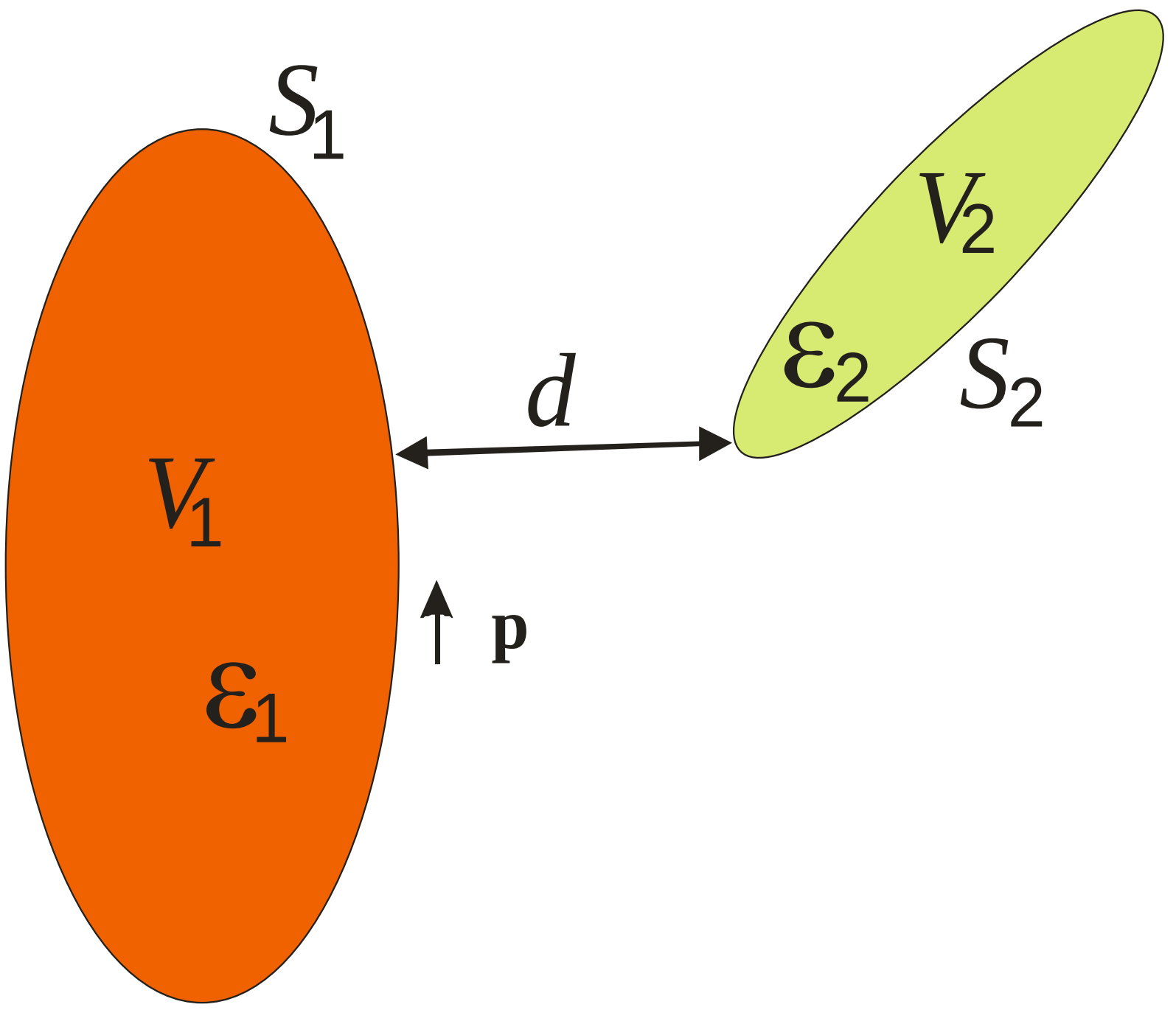

Fig. 1. Two arbitrary dielectric bodies excited by a dipole **p**

In the case of homogeneous and isotropic bodies of arbitrary shape, we consider DP described by a function $\varepsilon(\omega) = \varepsilon'(\omega) - i\varepsilon''(\omega)$ at temperatures $T_1$ and $T_2$, respectively, in a thermal equilibrium Planck field (large thermostat) with a temperature of $T_0$. We take the dependence on time in the form $\exp(i\omega t)$. Over time, the temperature of the bodies tends to equalize to $T_0$ and to achieve thermodynamic equilibrium with the thermal field. If the bodies are massive, and their heat capacity and thermal conductivity are large, then the temperatures can be considered constant. As a special case, we consider all temperatures to be equal to *T*. Let's consider this case first. Formally, thermal radiation should be taken into account in Maxwell's equations as an external field $\mathbf{E}_0$, $\mathbf{H}_0$. Zero vacuum fluctuations should also be included in this field. We will use the Rytov-Levin concept [8,56] to introduce into Maxwell's equations third-party fluctuation sources $\mathbf{J}_0(\omega,\mathbf{r},T) = i\omega\varepsilon_0(\varepsilon(\omega,\mathbf{r})-1)\mathbf{E}_0(\omega,\mathbf{r},T)$ related to the temperature of bodies, as well as with zero vacuum fluctuations, and outside the bodies $\mathbf{J}_0(\omega,\mathbf{r},0) = i\omega\mathbf{D}_0(\omega,\mathbf{r},0)$. If the bodies are very long, then far from their boundaries, the heat sources have zero average values $\langle \mathbf{J}_0(\omega,\mathbf{r},T)\rangle = 0$ and satisfy (in the SI system) the CS [8,56]

$$\left\langle J_{0\alpha}(\omega,\mathbf{r},T), J^*_{0\beta}(\omega,\mathbf{r}',T)\right\rangle = -2i\omega\varepsilon_0\Theta(\omega,T)\left[\varepsilon^*_{\alpha\beta}(\omega,\mathbf{r}) - \varepsilon_{\beta\alpha}(\omega,\mathbf{r})\right]\delta(\mathbf{r}-\mathbf{r}'). \tag{2}$$

To the left and right in (1) we have dimensions A$^2$×s/m$^4$. If DP is scalar, i.e. the bodies are isotropic (which we will use further for simplicity), then

$$\left\langle J_{0\alpha}(\omega,\mathbf{r},T), J^*_{0\beta}(\omega,\mathbf{r}',T)\right\rangle = 4\omega\varepsilon_0\varepsilon''(\omega,\mathbf{r})\Theta(\omega,T)\delta_{\alpha\beta}\delta(\mathbf{r}-\mathbf{r}'). \tag{2}$$

Let's write out similar CR for other quantities:

$$\left\langle E_{0\alpha}(\omega,\mathbf{r}), E_{0\beta}(\omega,\mathbf{r}')\right\rangle = 4\varepsilon''(\omega,\mathbf{r})\Theta(\omega,T)\delta_{\alpha\beta}\delta(\mathbf{r}-\mathbf{r}')/\left(\varepsilon_0|\varepsilon(\omega,\mathbf{r})-1|^2\omega\right),$$

$$\left\langle D_{0\alpha}(\omega,\mathbf{r},T), D^*_{0\beta}(\omega,\mathbf{r}',T)\right\rangle = 4\varepsilon_0\varepsilon''(\omega,\mathbf{r})\Theta(\omega,T)\delta_{\alpha\beta}\delta(\mathbf{r}-\mathbf{r}')/\omega.$$

We believe that DP does not depend on temperature. Here $\alpha,\beta = x,y,z$, and the function included in the relationship is the average energy of the quantum oscillator

$$\Theta(\omega,T) = \hbar\omega/2 + \hbar\omega f_{BE}(\omega,T), \tag{3}$$

expressed in terms of the Bose-Einstein function $f_{BE}(\omega,T) = \exp(\hbar\omega/k_bT) - 1$. Let the space be filled with an equilibrium thermal field. Such a field can also be considered as a third-party field. There is a Planck spectral radiation density $u(\omega,T) = f_{BE}(\omega,T)\hbar\omega^3/(\pi^2c^3)$, and the integral volumetric density is given by the integral of it over positive frequencies. It is equal to $U(T) = 4\sigma_{SB}T^4/c$, where $\sigma_{SB} = \pi^4k_B^4/(60c^2\hbar^3)$ is the Stefan-Boltzmann constant. The pressure of the thermal field is equal to $U(T)/3$. It acts on a single body from all sides, so it does not create a force. In addition to the thermal field, there is a field of zero vacuum fluctuations. They contain energies of all frequencies $\hbar\omega/2$ in standing vibrations that do not transfer energy. Away from the bodies they can be considered as waves of all directions $E^0_{0\alpha}(\omega,t) = E^0_{0\alpha}(\omega)\cos(\omega(t \mp x_\alpha/c) + \varphi^{\mp}_{\omega\alpha})$ with the same amplitudes and random phases $x_\alpha = (x,y,z)$, $\varphi_{\omega\alpha} = (\varphi_{\omega x}, \varphi_{\omega y}\varphi_{\alpha z})$. For such waves $H^0_{0\alpha}(\omega,t) = \sqrt{\varepsilon_0/\mu_0}E^0_{0\alpha}\cos(\omega(t \mp x_\alpha/c) + \varphi^{\mp}_{\omega\alpha})$

, $E_{0\alpha}(\omega)=E_0(\omega)$. The same waves occur in the thermal field. From the amplitude we denote as $E_{0\alpha}$. These waves can carry energy, unlike zero oscillations, and are created by distant heated bodies. For example, in a room with the same temperature in all points, including thick walls, they are created by random radiation from atoms in the walls. The walls also completely absorb the incident radiation, and emit the same amount, while their temperature is constant.

Consider dielectric bodies. They are described by integral equations in the form of LSIEs, having the form

$$\mathbf{E}(\omega,\mathbf{r},T)=\mathbf{E}^0(\omega,\mathbf{r},T)+\int\limits_{V=V_1+V_2}\hat{\Gamma}^{ee}(\omega,r-r')\mathbf{J}(\omega,\mathbf{r}',T)d^3r', \tag{4}$$

$$\mathbf{H}(\omega,\mathbf{r},T)=\mathbf{H}^0(\omega,\mathbf{r},T)+\int\limits_{V=V_1+V_2}\hat{\Gamma}^{he}(\omega,\mathbf{r}-\mathbf{r}')\mathbf{J}(\omega,\mathbf{r}',T)d^3r'. \tag{5}$$

Fields $\mathbf{E}^0(\omega,\mathbf{r},T)$, $\mathbf{H}^0(\omega,\mathbf{r},T)$ falling from infinity should be considered independent of $\mathbf{r}$. The bodies disturb the fields near their boundaries. Equation (4) becomes integral by virtue of $\mathbf{J}(\omega,\mathbf{r},T)=i\omega\varepsilon_0(\varepsilon(\omega)-1)\mathbf{E}(\omega,\mathbf{r},T)$. After obtaining solution of equation (4), we can find the magnetic field from (5). Fields (4), (5) have zero average values $\langle\mathbf{E}(\omega,\mathbf{r},T)\rangle=0$, but non-zero correlations. These correlations $\langle\mathbf{E}(\omega,\mathbf{r},T),\mathbf{E}^*(\omega',\mathbf{r}',T)\rangle$, $\langle\mathbf{H}(\omega,\mathbf{r},T),\mathbf{H}^*(\omega',\mathbf{r}',T)\rangle$ must be found. Correlations are understood as Gibbs ensemble averages $\langle\mathbf{E}(\omega,\mathbf{r},T),\mathbf{E}^*(\omega',\mathbf{r}',T)\rangle=\left(\mathbf{E}(\omega,\mathbf{r},T)\cdot\mathbf{E}^*(\omega,\mathbf{r}',T)\right)_{\text{avar}}$. For a random function of time with zero average value $\langle f(t)\rangle=0$ in the stationary process its correlation $\langle f(t),f(t')\rangle=\langle f(t),f(t)\rangle\delta(t-t')$, $\langle f(t),f(t)\rangle$ does not depend on time, since time averaging $\langle f(t),f(t')\rangle$ gives a dependence on : $\tau=t-t'$: $\langle f(t),f(t)\rangle=\langle f(t'),f(t')\rangle$.. For GF, we have spectral representations [57]

$$\hat{\Gamma}^{ee}_{\alpha\beta}(\omega,r)=\frac{1}{(2\pi)^3}\int\limits_{-\infty}^{\infty}\hat{g}^{ee}_{\alpha\beta}(\omega,\mathbf{k})\exp(-i\mathbf{kr})d^3k, \tag{6}$$

$$\hat{\Gamma}^{he}_{\alpha\beta}(\omega,r)=\frac{1}{(2\pi)^3}\int\limits_{-\infty}^{\infty}\hat{g}^{he}_{\alpha\beta}(\omega,\mathbf{k})\exp(-i\mathbf{kr})d^3k, \tag{7}$$

$$\hat{g}^{ee}_{\alpha\beta}(\omega,\mathbf{k})=\frac{i}{\omega\varepsilon_0\left(k_0^2-k^2\right)}\begin{bmatrix}k_0^2-k_x^2 & -k_xk_y & -k_xk_z\\ -k_xk_y & k_0^2-k_y^2 & -k_yk_z\\ -k_xk_z & -k_yk_z & k_0^2-k_z^2\end{bmatrix}, \tag{8}$$

$$\hat{g}^{he}_{\alpha\beta}(\omega,\mathbf{k})=\frac{-i}{k_0^2-k^2}\begin{bmatrix}0 & k_z & -k_y\\ -k_z & 0 & k_x\\ k_y & -k_x & 0\end{bmatrix}, \tag{9}$$

$k^2 = k_x^2 + k_y^2 + k_z^2$, $\hat{g}^{ee}_{\alpha\beta}(\omega,\mathbf{k}) = \hat{g}^{ee}_{\beta\alpha}(\omega,\mathbf{k}) = -\hat{g}^{ee*}_{\beta\alpha}(\omega,\mathbf{k})$, $\hat{g}^{he}_{\alpha\beta}(\omega,\mathbf{k}) = -\hat{g}^{he}_{\beta\alpha}(\omega,\mathbf{k}) = \hat{g}^{he*}_{\beta\alpha}(\omega,\mathbf{k})$. To determine the force acting on the first body, it is necessary to determine the correlations on its boundary $\left\langle E_\alpha(\omega,\mathbf{r},T), E^*_\alpha(\omega,\mathbf{r},T)\right\rangle_{\mathbf{r}\in S_1}$, $\left\langle H_\alpha(\omega,\mathbf{r},T), H^*_\alpha(\omega,\mathbf{r},T)\right\rangle_{\mathbf{r}\in S_1}$. They are included in the Maxwell stress tensor with components

$$\begin{aligned}\hat{T}_{\alpha\beta}(t,\mathbf{r},T) &= \varepsilon_0\left\langle E_\alpha(t,\mathbf{r},T), E^*_\beta(t,\mathbf{r},T)\right\rangle + \mu_0\left\langle H_\alpha(t,\mathbf{r},T), H^*_\beta(t,\mathbf{r},T)\right\rangle - \\ &-\frac{1}{2}\delta_{\alpha\beta}\left[\varepsilon_0\left\langle E_\gamma(t,\mathbf{r},T), E^*_\alpha(t,\mathbf{r},T)\right\rangle + \mu_0\left\langle H_\gamma(t,\mathbf{r},T), H^*_\alpha(t,\mathbf{r},T)\right\rangle\right]\end{aligned} . \qquad (10)$$

Correlations are taken here, since the fields are random, and the summation is based on the index $\gamma=x,y,z$. Formally, the time is indicated, but the correlations do not depend on it, i.e. all the values in (10) are correlation coefficients. As for a random function of time $\langle f(t), f(t')\rangle = C(t-t')$ is a correlation function, then $\langle f(t), f(t)\rangle = C(0)$ is the correlation coefficient. We have [8]

$$\begin{aligned}&\langle f(t), f(t')\rangle = C(t-t') = \\ &= \frac{1}{(2\pi)^2}\int_{-\infty}^{\infty}\int_{-\infty}^{\infty}\langle f(\omega), f(\omega')\rangle \exp(i\omega t' - i\omega' t)\,d\omega\, d\omega' ,\\ &\langle f(\omega), f(\omega')\rangle = C(\omega)\delta(\omega-\omega').\end{aligned} \qquad (11)$$

Here $C(\omega) = \langle f(\omega), f(\omega)\rangle$ is the spectral power density (spectral intensity), which is an even function. We have

$$\begin{aligned}C(t-t') &= \frac{1}{2\pi}\int_{-\infty}^{\infty}\langle f(\omega), f(\omega)\rangle \exp(i\omega(t-t'))\,d\omega = \\ &= \frac{1}{\pi}\int_0^{\infty}\langle f(\omega), f(\omega)\rangle \cos(\omega(t-t'))\,d\omega\end{aligned} . \qquad (12)$$

Since the processes are stationary in time and correlations are independent of time, then

$$\begin{aligned}&\langle E_\alpha(t,\mathbf{r}), E_\alpha(t,\mathbf{r})\rangle = \langle E_\alpha(0,\mathbf{r}), E_\alpha(0,\mathbf{r})\rangle = \\ &= \frac{1}{\pi}\int_0^{\infty}\langle E_\alpha(\omega,\mathbf{r}), E_\alpha(\omega,\mathbf{r})\rangle \cos(\omega t)\,d\omega = \\ &= \frac{1}{\pi}\int_0^{\infty}\langle E_\alpha(\omega,\mathbf{r}), E_\alpha(\omega,\mathbf{r})\rangle\,d\omega\end{aligned} . \qquad (13)$$

In equation (13) $\alpha = x, y, z$. In some cases, by repeating indexes, we will understand the summation, which we will indicate. We have (summation by $\gamma$)

$$\hat{T}_{\alpha\beta}(\omega,\mathbf{r},T)=\frac{\varepsilon_0}{2}\left\langle E_\alpha(\omega,\mathbf{r},T),E^*_\beta(\omega,\mathbf{r},T)\right\rangle+\frac{\mu_0}{2}\left\langle H_\alpha(\omega,\mathbf{r},T),H^*_\beta(\omega,\mathbf{r},T)\right\rangle- \\ -\frac{1}{4}\delta_{\alpha\beta}\left[\varepsilon_0\left\langle E_\gamma(\omega,\mathbf{r}),E^*_\gamma(\omega,\mathbf{r})\right\rangle+\mu_0\left\langle H_\gamma(\omega,\mathbf{r}),H^*_\gamma(\omega,\mathbf{r})\right\rangle\right], \qquad (14)$$

$$\hat{T}_{\alpha\beta}(t,\mathbf{r},T)=\frac{1}{\pi}\int_0^\infty \hat{T}_{\alpha\beta}(\omega,\mathbf{r},T)d\omega. \qquad (15)$$

Let us write the relation (4) and its conjugate for the component $x$ in the form

$$E_x(\omega,\mathbf{r},T)=E^0_x(\omega,T)+i\omega\varepsilon_0(\varepsilon(\omega)-1)\int_{V=V_1+V_2}\hat{\Gamma}^{ee}_{x\alpha}(\omega,\mathbf{r}-\mathbf{r}')E_\alpha(\omega,\mathbf{r}',T)d^3r',$$

$$E^*_x(\omega,\mathbf{r},T)=E^0_x(\omega,T)-i\omega\varepsilon_0(\varepsilon^*(\omega)-1)\int_{V=V_1+V_2}\hat{\Gamma}^{ee*}_{x\alpha}(\omega,\mathbf{r}-\mathbf{r}')E_\alpha(\omega,\mathbf{r}',T)d^3r'.$$

The value $E^0_x(\omega,T)$ is real. To do the correlations, we have

$$\left\langle E_x(\omega,\mathbf{r},T),E^*_x(\omega,\mathbf{r},T)\right\rangle=\left\langle E^0_x(\omega,T),E^0_x(\omega,T)\right\rangle+ \\ +i\omega\varepsilon_0(\varepsilon(\omega)-1)\int_{V=V_1+V_2}\hat{\Gamma}^{ee}_{x\alpha}(\omega,\mathbf{r}-\mathbf{r}')\left\langle E_\alpha(\omega,\mathbf{r}',T),E^0_x(\omega,T)\right\rangle d^3r'- \\ -i\omega\varepsilon_0(\varepsilon^*(\omega)-1)\int_{V=V_1+V_2}\hat{\Gamma}^{ee*}_{x\alpha}(\omega,\mathbf{r}-\mathbf{r}')\left\langle E^*_\alpha(\omega,\mathbf{r}',T),E^0_x(\omega,T)\right\rangle d^3r'+ \\ -(\omega\varepsilon_0)^2|\varepsilon(\omega)-1|^2\int_{V=V_1+V_2}\hat{\Gamma}^{ee}_{x\alpha}(\omega,\mathbf{r}-\mathbf{r}')\hat{\Gamma}^{ee*}_{\alpha x}(\omega,\mathbf{r}-\mathbf{r}')\left\langle E_\alpha(\omega,\mathbf{r}',T),E^*_\alpha(\omega,\mathbf{r}',T)\right\rangle d^3r'. \qquad (16)$$

Here, the summation is based on the variable $\alpha$. Due to the limited volume of $V$ in the far zone, the fields from the bodies are represented by outgoing spherical waves and decrease. It should be considered $\left\langle E_\alpha(\omega,\mathbf{r}',T),E^0_x(\omega,T)\right\rangle=0$. If there are no bodies, then $\left\langle E_x(\omega,\mathbf{r},T),E^*_x(\omega,\mathbf{r},T)\right\rangle=\left\langle E^0_x(\omega,T),E^0_x(\omega,T)\right\rangle$. This is a zero approximation. The first approximation is obtained by substituting the zero in the right part of the integral. The second approximation is obtained by substituting the first approximation into the right-hand side, etc. As a result, we get a series. If $\varepsilon(\omega)\approx 1$, then the first approximation is enough accurate, and then we have

$$\left\langle E_x(\omega,\mathbf{r},T),E^*_x(\omega,\mathbf{r},T)\right\rangle=\left\langle E^0_x(\omega,T),E^0_x(\omega,T)\right\rangle+ \\ +(\omega\varepsilon_0)^2|\varepsilon(\omega)-1|^2\left\langle E^0_\alpha(\omega,T),E^0_\alpha(\omega,T)\right\rangle\int_{V=V_1+V_2}\hat{\Gamma}^{ee}_{x\alpha}(\omega,\mathbf{r}-\mathbf{r}')\hat{\Gamma}^{ee*}_{x\alpha}(\omega,\mathbf{r}-\mathbf{r}')d^3r'. \qquad (17)$$

To be able to calculate (17), correlations should be determined. Since the value $\varepsilon_0\left\langle E^0_\alpha(\omega,T),E^0_\alpha(\omega,T)\right\rangle/2$ should be associated with the energy of electrical fluctuations, it should be proportional to $\Theta(\omega,T)$. Its dimension is J×s/m$^3$, therefore, it should be taken as $\varepsilon_0 \left\langle E^0_\alpha(\omega,T), E^0_\alpha(\omega,T)\right\rangle/2 = \Theta(\omega,T)\omega^2/(\pi^2 c^3)$, from where, due to the uniform distribution of energy over polarizations

$$\left\langle E^0_x(\omega,T), E^0_x(\omega,T)\right\rangle = 2\Theta(\omega,T)\omega^2/(3\varepsilon_0\pi^2 c^3). \tag{18}$$

At $T=0$ , we have $\Theta(\omega,0)=\hbar\omega/2$, and relation (18) takes into account zero fluctuations of the quantum vacuum. In particular $\left\langle E^0_x(\omega,0), E^0_x(\omega,0)\right\rangle = \hbar\omega^3/(3\varepsilon_0\pi^2 c^3)$. Then to a first approximation

$$\begin{aligned}\left\langle E_x(\omega,\mathbf{r},T), E^*_x(\omega,\mathbf{r},T)\right\rangle &= \frac{\hbar\omega^3}{3\varepsilon_0\pi^2 c^3} + \\ &+ \frac{\hbar\omega^5\varepsilon_0|\varepsilon(\omega)-1|^2}{3\pi^2 c^3}\int_{V=V_1+V_2}\hat{\Gamma}^{ee}_{x\alpha}(\omega,\mathbf{r}-\mathbf{r}')\hat{\Gamma}^{ee*}_{x\alpha}(\omega,\mathbf{r}-\mathbf{r}')d^3r'\end{aligned}. \tag{19}$$

When integrating (19), the first term diverges in frequency. However, it does not depend on the bodies and their configuration and is associated with the infinite energy of the quantum vacuum. To obtain the final solution, it should be omitted. At high frequencies $\hat{g}^{ee}_{\alpha\beta}(\omega,\mathbf{k}) \sim 1/\omega$, $\hat{\Gamma}^{ee}_{\alpha\beta}(\omega,\mathbf{r}) \sim 1/\omega$, $|\varepsilon(\omega)-1|^2 \sim 1/\omega^4$, therefore the right-hand side in (19) is proportional $1/\omega$. The limit value at a high frequency of GF has the form $\hat{g}^{ee}_{\alpha\beta}(\omega,\mathbf{k})_{\omega\to\infty} = i\delta_{\alpha\beta}/(\omega\varepsilon_0)$, i.e. it does not depend on **k**. Consequently, all the values obtained on its basis do not depend on the configurations of bodies and correspond to the infinite energy of the quantum vacuum. Therefore, it should be replaced $\hat{g}^{ee}_{\alpha\beta}(\omega,\mathbf{k}) \to \hat{g}^{ee}_{\alpha\beta}(\omega,\mathbf{k}) - i\delta_{\alpha\beta}/(\omega\varepsilon_0)$. With such a GF, the frequency integral converges as $1/\omega^2$. For magnetic field correlations, we have

$$\begin{aligned}\left\langle H_x(\omega,\mathbf{r},T), H^*_x(\omega,\mathbf{r},T)\right\rangle &= \frac{\hbar\omega^5\varepsilon_0|\varepsilon(\omega)-1|^2}{3\pi^2 c^3}\times \\ &\times \int_{V=V_1+V_2}\hat{\Gamma}^{he}_{x\alpha}(\omega,\mathbf{r}-\mathbf{r}')\hat{\Gamma}^{he*}_{x\alpha}(\omega,\mathbf{r}-\mathbf{r}')d^3r'\end{aligned}. \tag{20}$$

The right-hand side of (20) converges as $1/\omega^3$. You can also subtract the limit value of GF and get even better convergence. To calculate the frequency integrals with FG (8), (9) , we should replace [57] $k_0^2 \to k_0^2 + i\delta k^2$, where $\delta k^2$ is a small positive value and understand the integrals as taking the real part from the limit at $\delta k^2 \to 0$. An alternative way to calculate correlations is to represent the tensor FGs in terms of a scalar function $G(r) = (4\pi r)^{-1}\exp(-i\omega r/c)$ and define the tensor FGS in the form $\hat{\Gamma}^{ee}_{\alpha\beta}(\omega,\mathbf{r}) = (i\omega\varepsilon_0)^{-1}(\delta_{\alpha\beta}k_0^2 + \partial_\alpha\partial_\beta)G(r)$, $\hat{\Gamma}^{he}_{\alpha\beta}(\mathbf{r}) = (\varepsilon_{\alpha\beta\gamma}\partial_\gamma)G(r)$. Here $\varepsilon_{\alpha\beta\gamma}$ is the Levi-Chivita tensor. This representation of GF is more convenient for calculating forces for bodies of arbitrary shape. Obviously, it is not difficult to write out all the other correlations included in the MST. The flow of the component $\hat{T}_{xx}$ through the surface of a body is the component $f_x$ of the force acting on the first body. The total force acting on the first body is

$$\mathbf{f}_1 = -\oint_{S_1} \hat{T}(t,\mathbf{r}_\perp)\mathbf{n}(\mathbf{r}_\perp) d^2 r \,. \tag{21}$$

It should be borne in mind that the shape of the bodies is taken into account both when integrating over the volume in (19), (20), and when integrating over the surface in (21). In the case of inhomogeneous bodies, the value $|\varepsilon(\omega,\mathbf{r}')-1|^2$ is included under the integral. In the case of anisotropic bodies, one should simply use CR of type (2). In the case of magnetic properties, additional terms arise in LSIE related to magnetic polarization. However, it is possible to describe bodies only by tensor DP, but taking into account SD. Then the appropriate CR should be used [8,56]. Formula (21) is determined by the momentum conservation theorem [54,55]. This theorem states that any field source emits a pulse, while it is affected by a reaction force equal to and opposite to the derivative of the radiated pulse in time. If the source emits a pulse equally in all opposite directions (a point dipole), the force does not act on it. If the source emits mainly in a dedicated direction (directional mirror antenna), then there is a recoil. Let the sources be tied to the body. A dielectric body can be considered as a set of point sources associated with it. The radiation of the body into the far zone in opposite directions is the same. If we have two bodies, then considering them as a whole, we also get that their radiation to the far zone is the same in opposite directions, i.e. there is no force on the totality of bodies as a whole. However, this does not mean that there is no interaction between the bodies or their parts. If the energy density between the bodies is less than that outside the bodies, then an attractive force acts. If this density is greater, there will be a repulsive force. In a system of two bodies plus a field, the law of conservation of momentum holds. The force is determined by the total inward flow of the Maxwell stress tensor through the full surface of the body. In this case, if the pulse is completely reflected (an ideal mirror), then the incoming pulse is doubled. The pressure on such a mirror is twice as much as on a fully absorbing screen. There may also be a field behind the mirror or screen, and then it should be taken into account. In this case, the energy density of zero vacuum fluctuations, which is infinite, should also be taken into account. Since the total momentum in the two-body system is conserved, the momentum transmitted to the first body is equal and opposite to the momentum transmitted to the second body. Therefore, the force acting on the first body is equal and opposite to the force acting on the second body. For the energy density of the thermal field, we have

$$U(T)=\frac{\varepsilon_0\langle \mathbf{E}(T),\mathbf{E}(T)\rangle}{2}+\frac{\mu_0\langle \mathbf{H}(T),\mathbf{H}(T)\rangle}{2}= \\ =3\varepsilon_0\int_0^\infty E_0^2(\omega)d\omega ,$$

$$u(\omega,T)=\frac{f_{BE}(\omega,T)\hbar\omega^3}{\pi^2c^3}=N_\omega\hbar\omega=3\varepsilon_0E_0^2(\omega,T).$$

This density exists far away from bodies because bodies emit spherical waves. It is finite because there are a finite number $N_\omega(T)$ of photons. The energy density of vacuum fluctuations $u(\omega,0)=N_\omega(0)\hbar\omega/2$ is infinite because the number of virtual photons is infinite. If the temperature of the bodies is different and different from the temperature of the thermal field, then the bodies at a great distance from each other either radiate or absorb heat. The fields radiated by the body are determined in the far zone by calculating the Poynting vector due to thermal sources determined by body temperature. In this case, the body absorbs radiation from the thermal field and the second body. It is possible to consider a constant body temperature only in the case of sufficiently extended bodies with significant heat capacity. In this case, the balanced heat conservation equations should be solved under the condition of quasi-equilibrium and the time to set the temperature to the radiation temperature should be determined. In this case, we consider radiation as a Langevin thermostat with infinite heat capacity.

### 3. Excitation of bodies by dipoles and correlations

Let there be an electric dipole with a moment **p** $=\mathbf{x}_0p_e$ at the point $\mathbf{r}_0$. The current density of the dipole has the form $\mathbf{J}(x,y,z)=\mathbf{p}_e\delta(x-x_0)\delta(y-y_0)\delta(z-z_0)$. It creates a vector potential

$$\mathbf{A}_{ex}(\omega,\mathbf{r})=\frac{\mathbf{x}_0p_e}{(2\pi)^3}\int_{-\infty}^{\infty}\frac{\exp(-i\mathbf{k}(\mathbf{r}-\mathbf{r}_0))}{k^2-k_0^2}d^3k= \\ =\frac{\mathbf{x}_0p_e}{4\pi|r-r_0|}\exp(-ik_0|r-r_0|) . \quad (22)$$

The electric field is determined by the formula

$$\mathbf{E}_{ex}^0(\omega,\mathbf{r})=\frac{k_0^2\hat{I}+\nabla\times\nabla}{i\omega\varepsilon_0}\mathbf{A}_e(\omega,\mathbf{r}). \quad (23)$$

In (23), the unit tensor is denoted. In the case of a magnetic dipole with a dipole moment with $\mathbf{m}=\mathbf{x}_0p_m$

$$\mathbf{A}_{mx}(\omega,\mathbf{r})=\frac{\mathbf{x}_0p_m}{(2\pi)^3}\int_{-\infty}^{\infty}\frac{\exp(-i\mathbf{k}(\mathbf{r}-\mathbf{r}_0))}{k^2-k_0^2}d^3k= \\ =\frac{\mathbf{x}_0p_m}{4\pi|r-r_0|}\exp(-ik_0|r-r_0|) , \quad (24)$$

$$\mathbf{E}^0_{mx}(\omega,\mathbf{r}) = \nabla \times \mathbf{A}_{mx}(\omega,\mathbf{r}). \tag{25}$$

It should be solved in the LSIE in the form of

$$\mathbf{E}_{(e,m)\alpha}(\omega,\mathbf{r}) = \mathbf{E}^0_{(e,m)\alpha}(\omega,\mathbf{r}) + \\ + i\omega\varepsilon_0(\varepsilon(\omega)-1) \int\limits_{V=V_1+V_2} \hat{\Gamma}^{ee}(\omega, r-r')\mathbf{E}_{(e,m)\alpha}(\omega,\mathbf{r}')d^3r' . \tag{26}$$

They should be solved inside the bodies, where the field of the dipole located outside the bodies is regular. Therefore, solutions are obtained in a standard way by decomposing the electric field and using the Galerkin method or the Ritz type variational method [58]. We have solutions $\mathbf{E}_{(e,m)\alpha}(\omega,\mathbf{r})$ where $\alpha = x, y, z$. In the absence of a thermal field, we have correlation relations in the form of [8,56]

$$\langle E_x(\mathbf{r}_1), E_x(\mathbf{r}_2)\rangle = -\Theta(\omega,T)\mathrm{Re}(\mathbf{x}_0\mathbf{E}_{ex}(\mathbf{r}_2))/\pi , \tag{27}$$

$$\langle H_x(\mathbf{r}_1), H_x(\mathbf{r}_2)\rangle = -\Theta(\omega,T)\mathrm{Re}(\mathbf{x}_0 H_{mx}(\mathbf{r}_2))/\pi . \tag{28}$$

In (27) $\mathbf{E}_{ex}(\mathbf{r}_2)$ is the field of excitation of the structure by an electric dipole located at a point $\mathbf{r}_2$. The point $\mathbf{r}_1$ corresponds to an electric dipole located along the x at this point. The dipole field itself is not taken into account in the field $\mathbf{E}_{ex}(\mathbf{r}_2)$. Similarly for (28), we have magnetic dipoles instead of electric ones. We find the field $\mathbf{H}_{mx}$ using the GF $\hat{\Gamma}^{he}(\omega, r-r')$. To determine the force, we must assume $\mathbf{r}_1 \in S_1$, $\mathbf{r}_2 = \mathbf{r}_1 + \mathbf{n}\delta$, where $\delta$ is a small value. Then all correlations are regular functions, and the integral over the surface in the case of a single body is zero. The simplest method considered solves the problem of the interaction of two flat sheets of current, for example, two sheets of graphene [59,60] or the Lifshitz problem [8]. In this case, the problem is solved by switching to a two-dimensional FG transformation, and algebraic equations with respect to two-dimensional Fourier transforms are obtained.

## 4. Determination of force by the Lorentz formula

Fluctuation currents in each of the two interacting bodies depend on the thermal field and radiation of the other body, i.e. fluctuations are proportional to $\Theta(\omega,T)$ and to functions depend on the configuration of the bodies, in particular, on the distance between them. We can describe bodies only in terms of fluctuating current densities $J_1(\omega,\mathbf{r},T)$, $J_2(\omega,\mathbf{r},T)$. The whole set of dependencies on body configurations is described by a real function $F_{nm}(\omega,\mathbf{r})$. Obviously $F_{nm}(\omega,\mathbf{r}) = F_{mn}(\omega,\mathbf{r}) = F(\omega,\mathbf{r})$, $F_{12}(\omega,\mathbf{r}) = 0$, $n,m = 1,2$. Then we have the CR

$$\left\langle J_{n\alpha}(\omega,\mathbf{r},T), J^*_{m\beta}(\omega,\mathbf{r}',T)\right\rangle = F(\omega,\mathbf{r})\omega\varepsilon_0\Theta(\omega,T)\delta_{\alpha\beta}\delta_{nm}\delta(\mathbf{r}-\mathbf{r}'). \tag{29}$$

Different polarizations are uncorrelated, as are currents on different bodies. Currents at different points are not correlated either. The fields of the first body have the form $\mathbf{E}_1(\omega,\mathbf{r},T)=\tilde{\Gamma}^{ee}_{V_1}(\mathbf{r},\mathbf{J}_1(\omega,\mathbf{r}',T))$, $\mathbf{H}_1(\omega,\mathbf{r},T)=\tilde{\Gamma}^{he}_{V_1}(\mathbf{r},\mathbf{J}_1(\omega,\mathbf{r}',T))$, where the integral operators acting on the current in the volume $V_1$ are indicated. The power radiated by the body has components

$$P_{1\alpha}(\omega,\mathbf{r},T)=\varepsilon_{\alpha\beta\gamma}\operatorname{Re}\left\langle E_{1\beta}(\omega,\mathbf{r},T),H^*_{1\gamma}(\omega,\mathbf{r},T)\right\rangle/2=$$
$$=\varepsilon_{\alpha\beta\gamma}\operatorname{Re}\int_{V_1}\int_{V_1}\hat{\Gamma}^{ee}_{\beta\delta}(\omega,\mathbf{r}-\mathbf{r}',T)\hat{\Gamma}^{he*}_{\gamma\rho}(\omega,\mathbf{r}-\mathbf{r}'',T)\left\langle J_{1\delta}(\omega,\mathbf{r}',T)J^*_{1\rho}(\omega,\mathbf{r}'',T)\right\rangle d^3r'd^3r'' . \quad (30)$$

There is a summation based on repeated indexes. The right-hand side (30) is a real value, and it is not necessary to take the real part. Given the delta correlation, we get

$$P_{1\alpha}(\omega,\mathbf{r},T)=\omega\varepsilon_0\Theta(\omega,T)\varepsilon_{\alpha\beta\gamma}\int_{V_1}\hat{\Gamma}^{ee}_{\beta\delta}(\omega,\mathbf{r}-\mathbf{r}',T)\hat{\Gamma}^{he*}_{\gamma\delta}(\omega,\mathbf{r}-\mathbf{r}',T)F(\omega,\mathbf{r}')d^3r' .$$

It is now possible to determine the total radiation power $\mathbf{P}(\omega,\mathbf{r},T)=\mathbf{P}_1(\omega,\mathbf{r},T)+\mathbf{P}_2(\omega,\mathbf{r},T)$ and its total flux through the distant spherical surface surrounding the bodies $Q(\omega,T)=4\pi r^2 P_r(\omega,\mathbf{r},T)_{r\to\infty}$. Since the fields are emitted by sources in the form of a set of dipoles, they are in the far zone $P_r(\omega,\mathbf{r},T)_{r\to\infty}\sim 1/r^2$, so the result does not depend on *r*, but depends on the distribution of sources inside the bodies, i.e. on the function *F*. Consider two bodies as one volume $V=V_1+V_2$. Thermal radiation with density $u_0(\omega,T)=\Theta(\omega,T)\omega^2/(\pi^2c^3)=3\varepsilon_0E_0^2(\omega,T)$ falls on it from infinity. We designated it with the index 0 and took the function $\Theta$, taking into account the fact that these can be waves of zero vacuum oscillations. We believe that it passes through the bodies in all three directions in both directions. In this case, we do not consider reflected waves, since we have already taken them into account as radiated by bodies. According to the Kirchhoff principle, a body emits as much at each frequency as it absorbs. To determine the *F* functions, we apply this principle in our case for heat losses. For a wave of amplitude $E_0$, the volume loss density will be $\omega\varepsilon_0\varepsilon''(\omega)E_0^2(\omega,T)/2$. There is a loss density $\omega\varepsilon_0\varepsilon''(\omega)\Theta'(\omega,T)\omega^2/(6\pi^2c^3)$ for two opposite waves traveling in the direction $\pm x$. There is a total loss density of waves in all directions $\omega\varepsilon_0\varepsilon''(\omega)\Theta'(\omega,T)\omega^2/(2\pi^2c^3)$. Here, the stroke means that zero vacuum oscillations are not considered. Without vacuum fluctuations, take instead (29)

$$\left\langle J_{n\alpha}(\omega,\mathbf{r},T),J^*_{m\beta}(\omega,\mathbf{r}',T)\right\rangle=2F(\omega,\mathbf{r})\omega\varepsilon_0\varepsilon''(\omega)\Theta'(\omega,T)\delta_{\alpha\beta}\delta_{nm}\delta(\mathbf{r}-\mathbf{r}').$$

By equating the radiated and absorbed losses, we see that the value $\varepsilon''(\omega)$ decreases, and the equation for determining $F$ remains. To account for vacuum fluctuations, we omit the stroke in the final expressions.

Let consider the field of the second body acting on the first body. The first body has a surface charge density $\sigma(\omega,\mathbf{r}_\perp,T)=\varepsilon_0(1-1/\varepsilon_1(\omega))\mathbf{n}(\mathbf{r}_\perp)\mathbf{E}(\omega,\mathbf{r}_\perp,T)$ associated with a jump in the dielectric constant from $\varepsilon_1(\omega,\mathbf{r})$ to unity in vacuum. For the surface force density, take

$$\mathbf{f}_\sigma(\omega,\mathbf{r}_\perp,T)=\mathrm{Re}\left\langle\sigma(\omega,\mathbf{r}_\perp,T),\mathbf{E}_2^*(\omega,\mathbf{r}_\perp,T)\right\rangle/2. \tag{31}$$

In (31), the total field is used to determine the charge density, whereas only the electric field of the second body acts on the first body. The Lorentz force of the second field acts on the current density $\mathbf{J}_1(\omega,\mathbf{r},T)$ of first body. Its volumetric density is

$$f_{J\alpha}(\omega,\mathbf{r},T)=\varepsilon_{\alpha\beta\gamma}\mu_0\,\mathrm{Re}\left\langle J_{1\beta}(\omega,\mathbf{r},T),H_{2\gamma}^*(\omega,\mathbf{r},T)\right\rangle/2. \tag{32}$$

The total spectral force acting on the first body is

$$\mathbf{f}_1(\omega,T)=\oint_{S_1}\mathbf{f}_\sigma(\omega,T)\mathbf{n}(\mathbf{r}_\perp)d^2r+\int_{V_1}\mathbf{f}_J(\omega,\mathbf{r},T)d^3r. \tag{33}$$

To find the full force, we should calculate the integral

$$\mathbf{f}_1(T)=\frac{1}{\pi}\int_0^\infty\mathbf{f}_1(\omega,T)d\omega. \tag{34}$$

If the space between the bodies is filled with a medium with DP $\tilde{\varepsilon}(\omega)$ and magnetic MP $\tilde{\mu}(\omega,T)$, then $\sigma(\omega,\mathbf{r}_\perp,T)=\varepsilon_0(1-\tilde{\varepsilon}(\omega)/\varepsilon_1(\omega))\mathbf{n}(\mathbf{r}_\perp)\mathbf{E}(\omega,\mathbf{r}_\perp,T)$, and in formula (32) should be replaced $\mu_0\to\mu_0\tilde{\mu}(\omega,T)$. At the same time, when determining GFs, substitutions should be made everywhere the replacements $\varepsilon_0\to\varepsilon_0\tilde{\varepsilon}(\omega)$, $\mu_0\to\mu_0\tilde{\mu}(\omega,T)$, i.e. the method is applicable to dispersing medium. If the body is inhomogeneous, volumetric charge densities arise inside it, proportional to the gradient of DP, because we have $\nabla\cdot\mathbf{J}(\omega,\mathbf{r},T)=i\omega\varepsilon_0\mathbf{E}(\omega,\mathbf{r},T)\cdot\nabla\varepsilon(\omega,\mathbf{r})+i\omega\varepsilon_0(\varepsilon(\omega,\mathbf{r})-1)\nabla\cdot\mathbf{E}(\omega,\mathbf{r},T)$. From this continuity equation, we find the volumetric charge density $\rho(\omega,\mathbf{r},T)=i\nabla\cdot\mathbf{J}(\omega,\mathbf{r},T)/\omega=\varepsilon_0\mathbf{E}(\omega,\mathbf{r},T)\cdot\nabla\varepsilon(\omega,\mathbf{r})/\varepsilon(\omega,\mathbf{r})$.

## 5. Conclusions

Casimir forces or dispersion forces are considered for the interaction of two dielectric bodies of arbitrary configuration. Obtaining explicit ratios requires the introduction of a specific body configuration. The simplest results can be obtained for flat screens (for example, graphene sheets), or flat layers of finite thickness. In this case, the CRs depend on the distance between them and on the thickness. However, this needs to be considered in a separate paper. The results of the work can be easily generalized to heterogeneous and anisotropic structures. In this paper, the interactions of bodies at different temperatures are also not considered in detail. For some highly heated structures, when interacting in the far zone, vacuum fluctuations may have a weak effect compared to thermal ones, since the latter decrease more slowly with distance. Such an example is represented by carbon nanotubes [61,62]. Their radius of about 1 nm is several orders of magnitude smaller than the distance, so the boundary conditions change slightly, and thermal fluctuations prevail in determining the force.

**Financing the work**

The work was carried out with the financial support of the Ministry of Education and Science of the Russian Federation within the framework of the state assignment (project No. FSRR-2026-0006).